\documentclass[format=sigconf]{acmart}
\usepackage[export]{adjustbox}
\usepackage{url}
\usepackage{subfigure}
\usepackage{mathrsfs}

\AtBeginDocument{%
  \providecommand\BibTeX{{%
    \normalfont B\kern-0.5em{\scshape i\kern-0.25em b}\kern-0.8em\TeX}}}

\acmConference{}
\acmBooktitle{}
\acmSubmissionID{}
\acmYear{}
\acmDOI{}
\acmPrice{}

\begin{document}

\title{Notes on Refactoring Exponential Macros in Common Lisp\\ {\smaller Or: Multiple {\tt @Body} Considered Harmful}}

\author{Michael Wessel}
\email{michael.wessel@sri.com}
\affiliation{%
 \institution{SRI International}
 \streetaddress{333 Ravenswood Avenue}
  \city{Menlo Park}
  \state{California}
  \country{USA}
 \postcode{CA 94025}
}

\renewcommand{\shortauthors}{Michael Wessel}

\begin{abstract} 

  I recently consulted for a very big Common Lisp project having more
  than one million lines of code (including comments).  Let's call it
  ``System X'' in the following.  System X suffered from extremely
  long compilation times; i.e., a full recompile took about 33:17
  minutes on a 3.1 GHz MacBook Pro Intel Core i7 with SSD and 16 GBs
  of RAM, using ACL 10.1. It turns out that a number of macros were
  causing an exponential code blowup. With these macros refactored,
  the system then recompiled in 5:30 minutes --- a speedup by a factor
  of $\approx$ 6. In this experience report, I will first illuminate
  the problem, and then demonstrate two potential solutions in terms
  of macro refactoring techniques. These techniques can be applied in
  related scenarios. 
\end{abstract}

\keywords{Common Lisp, Macros, Exponential Code Blowup, Macro Refactoring, Very Large Lisp Systems}

\fancyfoot{} 

\settopmatter{printacmref=false} 
\renewcommand\footnotetextcopyrightpermission[1]{} 

\pagestyle{plain}

\settopmatter{printfolios=true}
\maketitle

\section{Introduction}

Macros and the ability to program language extensions in the language
\emph{itself} is one of the most beloved and powerful features of many
members of the Lisp family, and especially in Common Lisp
\cite{steele}, which has been coined a \emph{programmable programming
  language} by John Foderaro. The availability of the full programming
language at macro expansion / compile time makes Common Lisp an ideal
implementation platform for Domain Specific Languages
\cite{wessel2022,wessel2008}, and always has been (i.e., Lisp was an
early target platform for object-oriented programming concepts
\cite{loops}). Unlike macros in most other programming languages,
Common Lisp allows macros to be defined in the same language. Thanks
to its \emph{homoiconicity}, it offers a unified ``programs as data''
representation and allows the construction, manipulation, and most
importantly, \emph{computation} of macro expansions in the language
\emph{itself.} The full power of the language is always available --
not only at runtime, but also at \emph{macro-expansion (``compile'')
  time} \cite{graham1, graham2, norvig}.

As always, with great power comes great responsibility: macros can be
a double-edged sword. This is especially true in languages like Common
Lisp, where the main development mode is not the traditional ``edit
--- full recompile -- debug'' cycle, but an interactive, dynamic one,
based on incremental redefinition, evaluation, and
compilation. Unintended consequences of changes to the code base,
especially macros, can sometimes be left unnoticed for a longer time
period if full recompiles of the system are delayed. This holds true
especially in larger projects with bigger teams. Once compilation
times exceed half an hour, full recompilation is avoided by the
developers during daily development, and a build system will usually
be entrusted to deliver new base images overnight, containing the
changes of multiple developers. Of course, regressions will be
recorded and monitored on a daily basis. But even if \emph{build
  times} and \emph{the size of the fast load (FASL) files} are
reported by the build system, it might not be entirely clear which
changes increased the build time --- after all, the build system might
just have had a bad night and was busy performing backups as well, and
so on and so forth.

Consequently, tracing back unintended system behavior to (no longer so
recent) changes to the code base can become more difficult. For this
reason, incremental compilation of Common Lisp code can become a
drawback. I advise that developers should not only check for
unintended changes in semantics and functional characteristics of the
system caused by their code changes, but also to the non-functional
characteristics (e.g., FASL sizes and build time).  And especially for
macros.

\begin{figure*}[t!]
  \subfigure{\includegraphics[width=.6\columnwidth,valign=t]{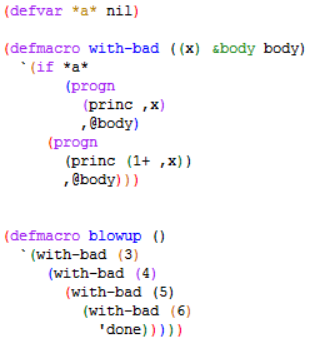}  } 
  \hspace*{1cm} \vrule \hspace*{1cm}
 \subfigure{\includegraphics[width=.95\columnwidth,valign=t]{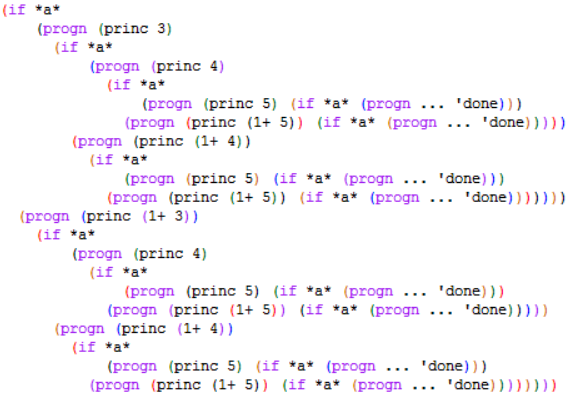}  } 
\vspace*{-1ex}\captionof{figure}{Macro with exponential macro expansion}
\label{fig:withbadexp} 
\end{figure*}

\begin{figure*}[t!]
  \subfigure{\includegraphics[width=.6\columnwidth,valign=t]{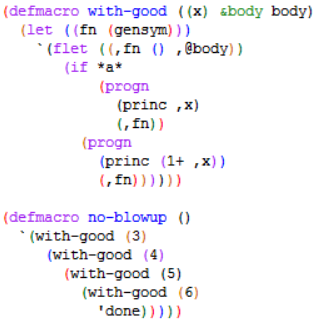}  } 
  \hspace*{1cm} \vrule \hspace*{1cm}
 \subfigure{\includegraphics[width=.95\columnwidth,valign=t]{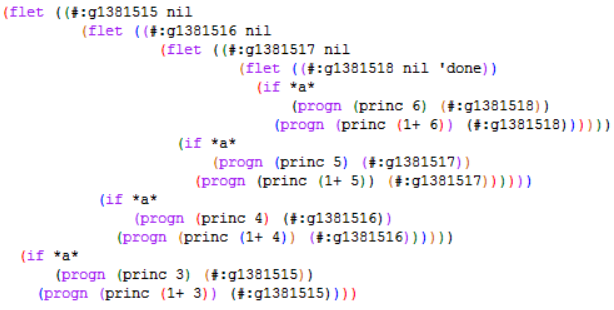}  } 
\vspace*{-1ex}\captionof{figure}{Macro with linear macro expansion}
\label{fig:withgoodexp} 
\end{figure*}

I recently had the opportunity to work on ``System X'', which is a
very large, multiple decades in-the-making Common Lisp system with
over one million lines of code.  System X suffered from extremely long
compilation times: a full compile required 33:17 minutes on a 3.1 GHz
MacBook Pro Intel Core i7 using an SSD and 16 GBs of RAM, using ACL
10.1. It turned out that \emph{three badly engineered macros} were
causing an exponential code blowup. With these macros refactored, a
full recompile is possible in 5:30 minutes --- a speedup by a factor
of $\approx$ 6. 

In this experience report, I will first illuminate the problem, and
then demonstrate two possible solutions in terms of macro
refactorings. The effectiveness of the refactoring methods is not only
demonstrated by the 6 fold reduction in compilation time, but also in
terms of FASL size reductions. The used methodology can be applied in
related scenarios. I conclude with some advice.

\section{The Problem }

The problems of System X are easily illustrated with a few synthetic
examples.  Consider the context-establishing {\tt with-bad} macro in
Figure \ref{fig:withbadexp}.  Like many {\tt with-} macros, it
utilizes a \emph{special variable} (here, {\tt *a*}) with dynamic
scope to control the runtime behavior beyond its lexical scope. In the
example, the binding of the dynamic variable {\tt *a*} determines if
{\tt with-bad} macro's expansion prints {\tt x} or {\tt (1+ x)}.

This macro serves to illustrate the problem of \emph{exponential macro
  expansion}. Frequently, {\tt with-} macros are nested, which can
obscure the magnitude of such problems from developers.  For example,
the macro might be part of a framework for website HTML generation and, 
as such, contain macros such as {\tt with-head}, {\tt with-body}, {\tt
  with-table}, and so on. Not only will complex web pages contain many
deeply nested occurrences of these macros, but it might also be the
case that certain common design elements of such pages (common
headers, footers, and navigation menus) have been aggregated into even
higher-level macros, which are then being used in other macros, and so
forth.

It should be noted that Common Lisp does not contain a \linebreak {\tt
  macroexpand-all} recursive macroexpansion facility; only \linebreak
{\tt macroexpand-1} is offered. This provides a single level of macro
expansion.  However, third-party solutions are available. I used one 
of these packages to diagnose the problems in System X \cite{arnesi, arnesi2}.

Considering the macroexpansion of {\tt blowup} containing four nested
{\tt with-bad} occurrences in Figure \ref{fig:withbadexp}, we can
clearly see that it is exponential in the size of the original
definition, due to the duplicated {\tt ,@body} forms. In general,
given a nesting depth of $n$, the size of the expanded macro code is
$2^n$. I even spotted {\tt with-} macros with more than two {\tt
  ,@body} forms ``in the wild''; in general, a {\tt with-} macro with
$m$ {\tt ,@body} forms will expand to size {$m^n$} if nested $n$
times. This should clearly be avoided.

Sometimes, such an \emph{exponential macro} is easy to fix. In the
case of Figure \ref{fig:withbadexp}, it suffices to move the {\tt
  ,@body} form into a local function definition ({\tt flet}) and call
the local function in the two places at runtime rather than
duplicating the code. The expansion size of the resulting {\tt
  with-good} macro shown in Figure \ref{fig:withgoodexp} is now linear
in the nesting depth rather than exponential.

Unfortunately, exponential macros are not always easily fixed. For
example, say the macro argument {\tt x} in {\tt with-good} was used to
establish lexical bindings for use \emph{within} the {\tt ,@body}
instead of just being an ``input parameter'' to the macro. In this
case, a {\tt (let ((,x ...)) ... ,@body)} would be used within the
macro to establish a corresponding lexical scope for {\tt
  ,x}. Moreover, the concrete binding to {\tt ,x} might depend on
complex runtime and compile time conditions. In particular, the value
of {\tt ,x} might depend on the runtime value of {\tt *a*}, which is
unknown at compile time / macroexpansion time, and hence, cannot be
anticipated by means of code rewritings / transformations. It is
thus important that the correct lexical contexts are established, 
for example, via the local function's lambdalist. 

\begin{figure}[t!]
  \centering
  \subfigure{\includegraphics[width=.9\linewidth]{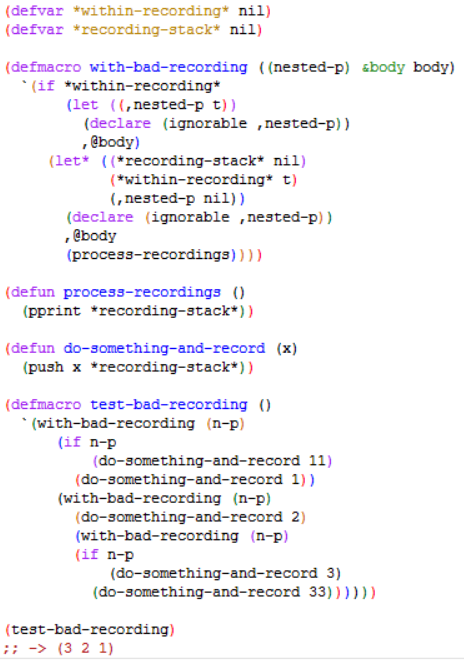} }
  \vspace*{-1ex}\captionof{figure}{The anti-pattern of an exponential
    {\tt with-} macro as found in System X}
\label{fig:badrecording} 
\end{figure}

Of such ``more difficult'' nature was the exponential macro that I had
to refactor in System X. Instead of revealing the details of this
macro I will use the synthetic example from Figure
\ref{fig:badrecording} in the following. This macro has a similar
complexity and serves to illustrate the problems and possible
solutions.

\noindent The idea behind {\tt with-bad-recording} is to establish a
context of dynamic scope for keeping track of ``instructions'' that
are being recorded onto a stack; these instructions can be entries to
a log file, an output recording presentation history, etc.  The
``hidden'' special variable {\tt *recording-stack*} (with dynamic
scope) is used to keep track of the values on the stack. This special
variable is not supposed to be visible to the user's code (it is
``internal''); instead, accessor functions (or macros) such as {\tt
  (do-something-and-record x)} are used to work with it.

Moreover, for whatever reason, clients of {\tt with-bad-recording}
also like to know whether the current invocation is toplevel, or
already part of a ``nested'' invocation at runtime; hence, a variable
{\tt nested-p} can be passed in which is then bound to {\tt nil} or
{\tt t}, respectively. To decide this, another special variable {\tt
  *within-recording*} had been (maybe redundantly) introduced by the
original author of the macro. Again, this is an ``internal'' special
variable which should not be visible to the user code, hence, {\tt
  nested-p} is supplied. A use case is shown in {\tt
  test-bad-recording}. Note that the runtime value of {\tt n-p} (i.e.,
{\tt ,nested-p} in the macro) is not knowable from the lexical
definition, as {\tt test-bad-recording} might occur nested within
another {\tt with-bad-recording} context at runtime. The use case
shows that the stack holds {\tt (3 2 1)} in the end.

\begin{figure}[t!]
  \centering
 \subfigure{\includegraphics[width=\linewidth]{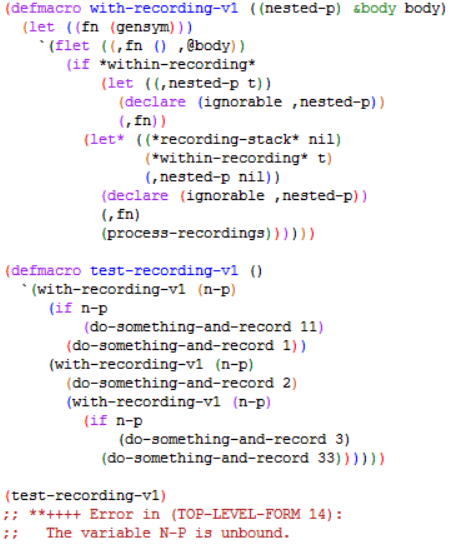}  } 
\vspace*{-1ex}\captionof{figure}{Replacing multiple {\tt ,@body} with local function calls broke the macro --- see the compiler warning}\vspace*{-3ex}
\label{fig:attemptv1} 
\end{figure} 

\noindent Clearly, this macro now has the potential for an exponential
macro expansion, and {\tt test-bad-recording} already suffers from
this blowup.  Can we fix this macro in the same way as in Figure
\ref{fig:withgoodexp}?

\section{First Solution --- Refactoring with {\tt FLET}}

\begin{figure}[t!]
  \centering
 \subfigure{\includegraphics[width=\linewidth]{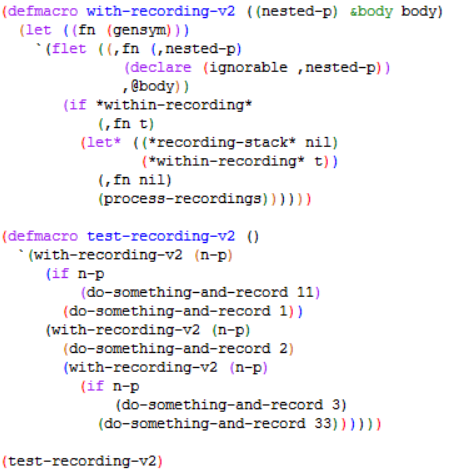}  } 
\vspace*{-1ex}\captionof{figure}{Using local function variables to establish lexical context --- note that the dynamic context is still established by the original branching structure}\vspace*{-3ex}
\label{fig:attemptv2} 
\end{figure}

The first solution is a generalization of the idea we already
discussed in Section 2 --- we replaced each {\tt ,@body} by a call to
a single local function containing a single {\tt ,@body}. 
We can use {\tt FLET} or {\tt LABELS} for that. 

A naive attempt of fixing {\tt with-bad-recording} is shown in Figure
\ref{fig:attemptv1}. This macro is now clearly broken, as {\tt ,@body}
refers to {\tt ,nested-p = n-p}, which is not visible in the outer
{\tt flet} - hence the compiler warning that this variable is now
unbound. The obvious solution is hence to make {\tt ,nested-p = n-p}
an argument of the local function so that the required lexical
variables for {\tt ,@body} are established by the local function. This
is shown in Figure \ref{fig:attemptv2}.

\begin{figure}[t!]
  \centering
  \subfigure{\includegraphics[width=\linewidth]{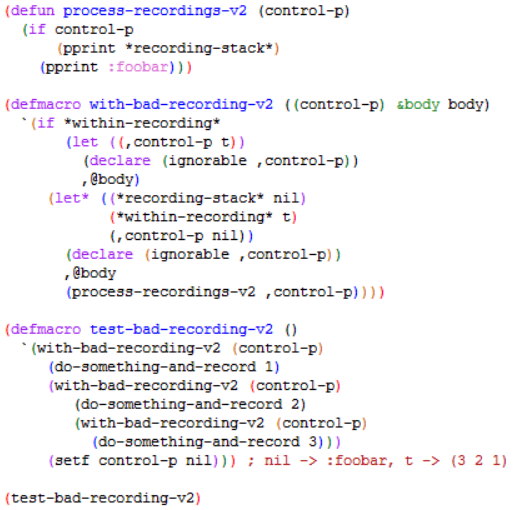} }
  \vspace*{-1ex}\captionof{figure}{If lexical variables that are
    arguments to the macro are modified, such as {\tt control-p}, then
    refactoring becomes more involved }\vspace*{-3ex}
\label{fig:badrecordingv2} 
\end{figure}

A further complication is introduced if the lexical variable is
modified in one of the branches --- consider the variation \linebreak
{\tt with-bad-recording-v2} shown in Figure \ref{fig:badrecordingv2},
where {\tt nested-p} is replaced by {\tt control-p}. The value of {\tt
  control-p} influences the output, and it might be set from either
within the user-supplied {\tt ,@body} code, or from within the macro
itself. Refactoring such a macro then becomes less mechanical, and more
care is needed to ensure that the right lexical environments are
established.

\noindent In particular, we realize that the value of {\tt control-p}
must be changed from \emph{within} the local function so that the call
\linebreak {\tt (process-recordings-v2 ,control-p)} will get the value
of \linebreak {\tt ,control-p} from the correct lexical scope. A
possible solution is shown in Figure \ref{fig:badrecordingv2}; the
{\tt branch-p} argument is used to inform the local function about the
invocation context, and, based on its value, the local function has to
accommodate, or ``emulate'', the different runtime behaviors from the
macro's original branches. This might not always be possible, but is
rather trivial in this case.\footnote{Please note that we are ignoring
  potential differences in the returned value of these macros for now;
  usually, {\tt with-} macros do not return values, but this is a
  convention and not a strict requirement.}

\begin{figure}[t!]
  \centering
 \subfigure{\includegraphics[width=\linewidth]{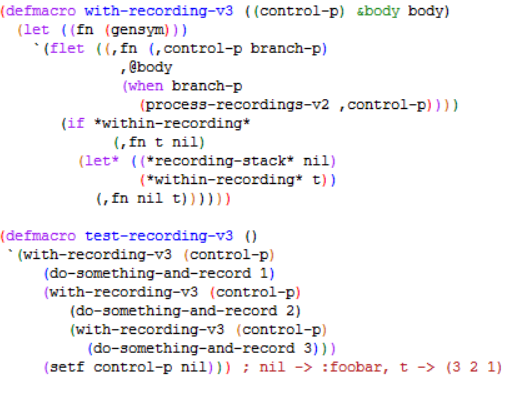}  } 
\vspace*{-1ex}\captionof{figure}{Accommodating different macro branches in one function with the {\tt branch-p} argument}\vspace*{-3ex}
\label{fig:attemptv3} 
\end{figure}

This wraps up the discussion of the first refactoring strategy.  In a
nutshell, the original branching structure establishing different
lexical and dynamic scopes is maintained. A common {\tt ,@body} form
must be found which is able to reproduce the original runtime
behaviors, and it is placed within a local function. The required
lexical contexts are established by the local function and removed
from the original branching structure. Additional control parameters
such as {\tt branch-p} are used to select branch-specific runtime
behavior. In particular, we maintain the branching structure of the
original macro in order to establish the right bindings for the
special variables, and to set up the correct lexical contexts by
calling the local function accordingly.

A potential drawback of the refactoring pattern just discussed is the
introduction of additional local functions and the additional runtime
overhead of additional function calls.\footnote{It might be possible to
  declare these local functions as {\tt inline} though.}

More severely, the (full) macro expansion of the refactored macro now
obfuscates the original structure of the macro --- it is ``inside
out'' because we employed \emph{functional composition} to implement
\emph{program sequencing;} as can be seen in Figure
\ref{fig:withgoodexp}, {\tt (princ 6)} now textually precedes {\tt
  (princ 3)}, contrary to the original definition.  In particular, the
{\tt ,@body}'s of the local functions are now ``detached'' from the
original branching structure, making the macro expansion more
difficult to understand. 

If these are serious concerns, the following alternative refactoring
strategy can be applied.

\section{Second Solution --- Refactoring with {\tt PROGV}}

In the following, we are \emph{not} using a local function that can be
called from different branches of the rewritten macro.  Instead we are
trying to unify the original branching structure establishing
different lexical and dynamic contexts into \emph{one} common
structure. It turns out that establishing the right (conditional)
bindings for the special variables is the biggest obstacle, and we
will be using {\tt progv} for this purpose.

\begin{figure}[t!]
  \centering
 \subfigure{\includegraphics[width=\linewidth]{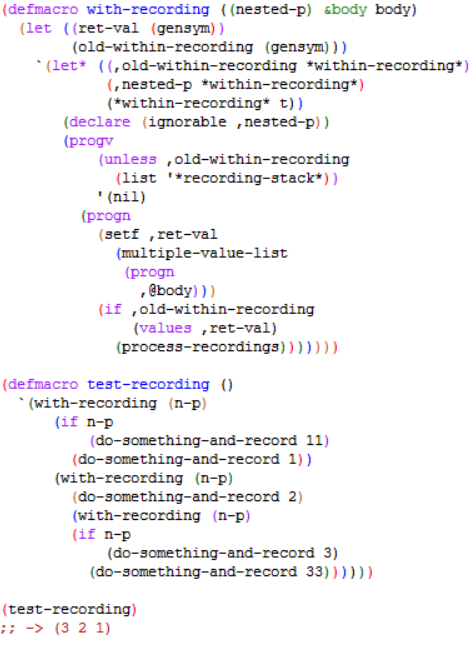}  } 
\vspace*{-1ex}\captionof{figure}{The rewritten linear {\tt with-} macro --- thanks to {\tt progv}}
\label{fig:recording} 
\end{figure}

The following set of steps can be understood as a semantics-preserving
code rewriting procedure / transformer.  We will apply the following
to the {\tt with-bad-recording} macro to tame the exponential beast
and rewrite it into a linear macro:

\begin{description}
\item[Step 1] Macroexpand / rewrite all branching special forms ({\tt
    unless}, {\tt when}, {cond}, \ldots) into {\tt if}s (in our 
  examples, this is already the case, so the step doesn't apply): 

\begin{verbatim}
(if <condition> 
     (let ( <binding 11> ... <binding 1n> ) 
         ,@body)
   (let ( <binding 21>  ... <binding 2m> )
       ,@body))   
\end{verbatim} 

\item[Step 2] Ensure that all {\tt let}s in all branches refer to the
  same variables, and in the same order. If {\tt <binding ij> = (,var
    val)} and {\tt var} is a macro argument, then all branches already
  must contain a valid {\tt (,var val)} binding. Otherwise, {\tt var} would
  be unbound in (some branches of) {\tt ,@body} (e.g., the macro was
  already defective in the first place).

  \quad If {\tt var} is a special variable instead, i.e., {\tt *var*}, then,
  in case the branch \emph{did not} contain a {\tt <binding> = (*var*
    val)}, we introduce a ``dummy'' binding {\tt <binding> =} \linebreak
  {\tt (*var* *var*)} for now.  The idea is to express that we \emph{intend}
  to \emph{not} alter the binding of {\tt *var*} dynamically.  Note
  that this is \emph{unproblematic} where {\tt *var*} is used as a
  ``read only'' variable, but problematic in cases such as {\tt
    with-bad-recording}, where {\tt *recording-} {\tt stack*} is
  modified; see below for the solution.

  \quad Hence, we now have the same number $k$ of {\tt (var val)} bindings
  in each {\tt let}, with potentially (not necessarily) different {\tt
    val}'s; note that $max(n,m) \leq k \leq n+m$:

\begin{verbatim}
(if <condition> 
     (let ( (<var1 val11>) ... (<vark val1k>) ) 
         ,@body)
  (let ( (<var1 val21>)  ... (<vark val2k>) )
      ,@body)) 
\end{verbatim} 

\item[Step 3] Next, we remove the different branches, establish all
  the bindings in a single {\tt let}, and recover the effects of the
  \linebreak {\tt <condition>} by establishing different bindings
  within the {\tt let} binding forms itself. Since we removed the
  different branches from the surrounding code by factoring in /
  moving the {\tt condition} \emph{into} the {\tt let} lambda lists,
  we have also eliminated the multiple {\tt ,@body} occurrences: 

\begin{verbatim}
(let ( ( <var1> (if <condition> <val11> <val21>) )
          ... 
          ( <vark> (if <condition> <val1k> <val2k>) ))
  ,@body) 
\end{verbatim}

\item[Step 4] So far so good --- there is one problem though: this 
  only works for dynamically scoped variables that are used in a
  ``read only'' fashion. As already mentioned, we have introduced a
  ``dummy'' binding {\tt <binding> =} \linebreak {\tt( *recording-stack*
    *recording-stack* )} to express that we wish to leave the binding of
  {\tt *recording-stack*} \emph{untouched.} But we changed it by
  establishing a new binding frame --- we ``shadowed'' the previously
  established binding.  With {\tt let/let*}, there is no solution to
  this.  

  \quad The effect is illustrated in Figure \ref{fig:attemptv4}
  --- the refactored macro is clearly broken now, as illustrated with
  the example call {\tt (test-recording-v4)}. Instead of returning
  {\tt (3 2 1)} like in the original, we are now only getting the
  first value that was pushed onto the stack: {\tt (1)}.

\begin{figure}[t!]
  \centering
 \subfigure{\includegraphics[width=.9\linewidth]{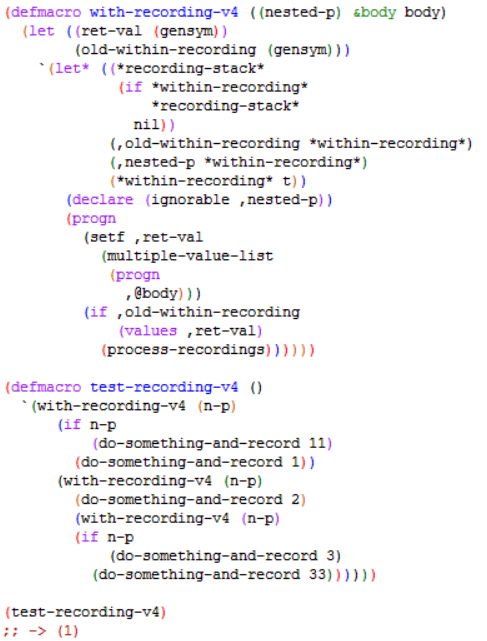}  } 
\vspace*{-1ex}\captionof{figure}{Shadowing {\tt *recording-stack*} broke the macro}
\label{fig:attemptv4} 
\end{figure}

\quad However, the code rewritings have brought us onto the right track.  We
only need to avoid shadowing in cases were we do not wish to alter a
dynamic variable. Fortunately, there is a solution to this in Common
Lisp, and one has to congratulate the designers of Common Lisp for
anticipating such a scenario: {\tt progv} can do the job as follows:

\begin{verbatim}
 (progv
   (when alters-*var*
       (list '*var*))
   (list val)
\end{verbatim} 

If {\tt alters-*var* = T}, the form is equivalent to {\tt (progv
  '(*var*) (list val))} hence establishing a new binding for {\tt
  *var*}. Otherwise, if {\tt alters-*var* = NIL}, then the form is
equivalent to {\tt (progv nil (list val))}, leaving {\tt *var*}
unchanged.

\quad Hence, the final step involves ``splitting up'' the single {\tt let}
(or {\tt let*}), and reestablishing the ``problematic'' special
bindings via {\tt progv} instead, in the manner just described. Since
this step is hard to templatize, let's look at the final rewritten
example macro in Figure \ref{fig:recording} instead. As can be seen
from the test invocation, it behaves correctly, and its expansion is
clearly linear.

\end{description}

\noindent Even though context establishing macros are usually not used
for their return values, it is nevertheless advisable to accommodate
for such, and so we did in Figure \ref{fig:recording}. Inspecting all
use cases of the macro in the source code of a very large system such
as System X to identify such use cases is more time consuming than to
cater for such cases correctly from the beginning. Hence, the
rewritten macro in Figure \ref{fig:recording} also returns the same
values as the original (utilizing {\tt multiple-value-list} and {\tt
  values}).

\section{Effectiveness of the Techniques}

We counted the number of macro function invocations (``macro expansion
calls'') that occurred during a full recompilation of System X and
compared the results between the original and the {\tt PROGV}
refactored versions.\footnote{Obviously, the {\tt FLET}-based
  technique will yield the same results.}

For the original version, we counted $12816 + 2431 + 2432 = 17679$
calls for our three critical exponential macros. Compared to
$882 + 530 + 531 = 1466$ invocations for the refactored version, with
a ratio of $17679 / 1466 = 12.05$.  Referring to the {$m^n$} notation
from Section 2, we have $m=2$ (two {\tt ,@body}'s), and can hence 
assume an average nesting depth of about $log_2 12.05 \approx 3.6 = n$.

\begin{figure}[t!]
  \centering
 \subfigure{\includegraphics[width=\linewidth]{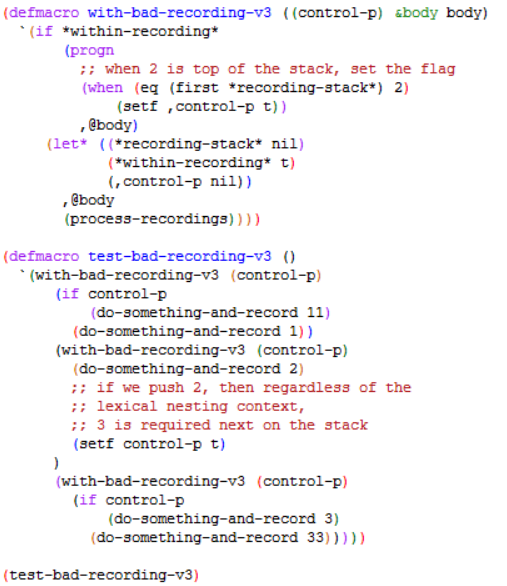}  } 
\vspace*{-1ex}\captionof{figure}{A macro that cannot be refactored using the discussed techniques}
\label{fig:problem} 
\end{figure}

The biggest FASL size reduction was observed for a file that shrank
from 53 MBs to only 2 MBs --- a factor of 26.5! Since
$log_2 26.5 \approx 4.7$ we can assume a more deeply nested use of the
exponential macros there.

These are rough estimates, but the numbers speak a clear language: for
large Lisp systems, the impact of even moderately deeply nested (i.e.,
$3 \leq n \leq 5$) exponential macros can be catastrophic in terms of
compilation time and FASL sizes.

\section{Limitations of the Techniques}

Whereas these refactoring patterns should cover a large region of
exponential {\tt with-} cases in practice, they are far from offering
a complete solution. An example of a macro that \emph{cannot} be
refactored with the so-far discussed techniques is shown in Figure
\ref{fig:problem}. Essentially, the problem is that the \emph{else}
branch establishes a lexical context for {\tt ,control-p}, but the
\emph{then} branch doesn't, and that it is impossible to know at
compile time which branch will be active. A solution akin to {\tt
  progv} would be needed, but for \emph{lexical} variables.

\section{Conclusion}

I presented techniques for refactoring exponential macros into linear
ones.  From my experience with System X, I learned that three (not
overly carefully designed) macros can suffice to severely (i.e.,
exponentially) affect compilation time and FASL size.

It is not entirely fair to blame the original designers of these
macros for causing so much trouble in the later life of System X. Each
project starts small, and the macros were originally doing fine. Only
later in System X's life cycle did the effects of exponential macro
expansion degrade its compilation time (and corresponding FASL file
sizes) drastically. Incremental development and compilation,
over-night build systems, multiple developers, and a focus on
\emph{functional} rather than \emph{non-functional} system
characteristics were all factors that contributed to code that grew
like a malignant cancer. The power of Common Lisp macros can be a
double-edged sword and needs to be handled with diligence and
delicacy.  Fortunately, Common Lisp is also powerful enough to offer a
cure to these problems, as we tried to illustrate in this Experience
Report. We hope that our experience will help other developers to
avoid such situations in their own projects.

Could some of this rewriting process be automated? For sure, compilers
could warn about potentially expensive macro expansions, or try to
identify exponential expansion.  A \linebreak {\tt macroexpand-all} as
part of a Common Lisp IDE would certainly help as well.
Interestingly, the code rewriting techniques described in Sections 3
and 4 seems straight-forward enough that it \emph{might be possible}
to automate, at least for certain macro patterns (but might be
undecidable in general). This could be interesting future research,
and I would appreciate any pointers and feedback from the Lisp
community --- surely, this problem is not new, yet I wasn't able to
find papers that would cover this topic. I hope that this report will
fill this gap, and also raise awareness in Common Lisp developers for
such issues.

\begin{acks}
  First and foremost, I would like to thank the Principial
  Investigator (PI) of Project X for allowing me to consult for System
  X, and for triggering the investigation of long compilation times in
  System X. This PI also read a Draft and suggested to substantiate
  the findings with hard numerical evidence. This led to the addition
  of Section 5. 

  Next, I wish to thank the anonymous reviewers for pointing out
  omissions in the original Draft that made the transition from the
  local function call-based solution (as illustrated with the {\tt
    with-good} macro) to the {\tt progv}-based solution seem
  unmotivated. As pointed out by the reviewers, the original Draft did
  mention the technique, but erroneously gave the impression that it
  could not be applied to accommodate for lexical variables. This is
  now demonstrated in Section 3, which was missing from the original
  Draft. I like to thank all reviewers for their comments and time.

  In my humble opinion, both refactoring methods have their own
  strengths and weaknesses, and I mentioned some of them at the end of
  Section 3.

\pagebreak

\noindent This work was funded by the National Institute of Allergy and
  Infectious Diseases of the National Institutes of Health under award
  number R01AI160719. The content is solely the responsibility of the
  authors and does not necessarily represent the official views of the
  National Institutes of Health. The NIH did not play any role in the
  design of the study; nor in collection, analysis, or interpretation
  of data; nor in writing the manuscript.
\end{acks}


\bibliographystyle{ACM-Reference-Format}
\bibliography{main}


\begin{thebibliography}{9}


\ifx \showCODEN    \undefined \def \showCODEN     #1{\unskip}     \fi
\ifx \showDOI      \undefined \def \showDOI       #1{#1}\fi
\ifx \showISBNx    \undefined \def \showISBNx     #1{\unskip}     \fi
\ifx \showISBNxiii \undefined \def \showISBNxiii  #1{\unskip}     \fi
\ifx \showISSN     \undefined \def \showISSN      #1{\unskip}     \fi
\ifx \showLCCN     \undefined \def \showLCCN      #1{\unskip}     \fi
\ifx \shownote     \undefined \def \shownote      #1{#1}          \fi
\ifx \showarticletitle \undefined \def \showarticletitle #1{#1}   \fi
\ifx \showURL      \undefined \def \showURL       {\relax}        \fi
\providecommand\bibfield[2]{#2}
\providecommand\bibinfo[2]{#2}
\providecommand\natexlab[1]{#1}
\providecommand\showeprint[2][]{arXiv:#2}

\bibitem[Baringer(2023a)]%
        {arnesi}
\bibfield{author}{\bibinfo{person}{Marco Baringer}.} \bibinfo{year}{Accessed:
  2-26-2023}\natexlab{a}.
\newblock \bibinfo{title}{{ARNESI Library}}.
\newblock
\newblock
\urldef\tempurl%
\url{https://bese.common-lisp.dev/arnesi.html}
\showURL{%
\tempurl}


\bibitem[Baringer(2023b)]%
        {arnesi2}
\bibfield{author}{\bibinfo{person}{Marco Baringer}.} \bibinfo{year}{Accessed:
  2-26-2023}\natexlab{b}.
\newblock \bibinfo{title}{{ARNESI Library}}.
\newblock
\newblock
\urldef\tempurl%
\url{https://quickref.common-lisp.net/arnesi.html}
\showURL{%
\tempurl}


\bibitem[Bobrow et~al\mbox{.}(1986)]%
        {loops}
\bibfield{author}{\bibinfo{person}{Daniel~G. Bobrow}, \bibinfo{person}{Kenneth
  Kahn}, \bibinfo{person}{Gregor Kiczales}, \bibinfo{person}{Larry Masinter},
  \bibinfo{person}{Mark Stefik}, {and} \bibinfo{person}{Frank Zdybel}.}
  \bibinfo{year}{1986}\natexlab{}.
\newblock \showarticletitle{CommonLoops: Merging Lisp and Object-Oriented
  Programming}. In \bibinfo{booktitle}{\emph{Conference Proceedings on
  Object-Oriented Programming Systems, Languages and Applications}} (Portland,
  Oregon, USA) \emph{(\bibinfo{series}{OOPSLA '86})}.
  \bibinfo{publisher}{Association for Computing Machinery},
  \bibinfo{address}{New York, NY, USA}, \bibinfo{pages}{17–29}.
\newblock
\showISBNx{0897912047}
\urldef\tempurl%
\url{https://doi.org/10.1145/28697.28700}
\showDOI{\tempurl}


\bibitem[Graham(1994)]%
        {graham2}
\bibfield{author}{\bibinfo{person}{P. Graham}.}
  \bibinfo{year}{1994}\natexlab{}.
\newblock \bibinfo{booktitle}{\emph{{On Lisp}}}.
\newblock \bibinfo{publisher}{Prentice-Hall}.
\newblock


\bibitem[Graham(1996)]%
        {graham1}
\bibfield{author}{\bibinfo{person}{P. Graham}.}
  \bibinfo{year}{1996}\natexlab{}.
\newblock \bibinfo{booktitle}{\emph{{ANSI Common Lisp}}}.
\newblock \bibinfo{publisher}{Prentice-Hall}.
\newblock


\bibitem[Michael~Wessel(2008)]%
        {wessel2008}
\bibfield{author}{\bibinfo{person}{Ralf~Möller Michael~Wessel}.}
  \bibinfo{year}{2008}\natexlab{}.
\newblock \showarticletitle{{ Software Abstractions for Description Logic
  Systems}}. In \bibinfo{booktitle}{\emph{Proceedings of the 5th European Lisp
  Workshop (ELW’08)}}.
\newblock
\urldef\tempurl%
\url{{http://www.european-lisp-workshop.org/archives/08.wessel.pdf}}
\showURL{%
\tempurl}


\bibitem[Norvig(1992)]%
        {norvig}
\bibfield{author}{\bibinfo{person}{P. Norvig}.}
  \bibinfo{year}{1992}\natexlab{}.
\newblock \bibinfo{booktitle}{\emph{{Artificial Intelligence Programming: Case
  Studies in Common Lisp}}}.
\newblock \bibinfo{publisher}{Morgan Kaufmann}.
\newblock


\bibitem[Steele(1990)]%
        {steele}
\bibfield{author}{\bibinfo{person}{Guy~L. Steele, Jr.}}
  \bibinfo{year}{1990}\natexlab{}.
\newblock \bibinfo{booktitle}{\emph{{Common Lisp -- The Language, Second
  Edition}}}.
\newblock \bibinfo{publisher}{Digital Press}.
\newblock


\bibitem[Wessel(2022)]%
        {wessel2022}
\bibfield{author}{\bibinfo{person}{Michael Wessel}.}
  \bibinfo{year}{2022}\natexlab{}.
\newblock \showarticletitle{{ An Ontology-Based Dialogue Management Framework
  for Virtual Personal Assistants in Common Lisp}}. In
  \bibinfo{booktitle}{\emph{European Lisp Symposium '23 (ELS'23)}}.
\newblock
\urldef\tempurl%
\url{{https://doi.org/10.5281/zenodo.6335631}}
\showURL{%
\tempurl}


\end{thebibliography}

\end{document}